\documentclass[conference,10pt]{IEEEtran}
\usepackage[T1]{fontenc}
\usepackage{amsmath,amssymb,booktabs,array,graphicx,tikz}
\usetikzlibrary{arrows.meta,positioning,fit,calc}
\usepackage{balance}
\usepackage{enumitem}
\newlist{contributions}{enumerate}{1}
\setlist[contributions]{label=\arabic*),leftmargin=1.7em,
  labelsep=.5em,topsep=3pt,partopsep=0pt,itemsep=3pt,parsep=0pt}
\usepackage[hidelinks]{hyperref}
\hypersetup{
  pdftitle={LocalTrain: Integer-Only Local Learning for Energy-Efficient CNN Training on Edge FPGAs},
  pdfauthor={Shiqi Ding, Xuanyu Liang, Xinyang Liu, Boyang Li, Zhiqiang Que, Guosheng Hu, Jiayang Li}
}
\definecolor{compute}{RGB}{233,240,248}
\definecolor{storage}{RGB}{242,242,242}
\definecolor{control}{RGB}{253,242,225}
\tikzset{box/.style={draw=black!70,fill=compute,minimum height=6mm,align=center,font=\footnotesize,inner sep=3pt},mem/.style={box,fill=storage},ctl/.style={box,fill=control},arr/.style={-{Latex[length=1.4mm]},line width=.45pt},dep/.style={arr,dashed},lab/.style={font=\scriptsize,align=center}}
\begin{document}
\title{LocalTrain: Integer-Only Local Learning for Energy-Efficient CNN Training on Edge FPGAs}
\author{
\IEEEauthorblockN{
Shiqi Ding\textsuperscript{1},
Xuanyu Liang\textsuperscript{2},
Xinyang Liu\textsuperscript{3},
Boyang Li\textsuperscript{1},
Zhiqiang Que\textsuperscript{3},
Guosheng Hu\textsuperscript{3},
Jiayang Li\textsuperscript{3*}
}
\IEEEauthorblockA{
\textsuperscript{1}University of Cambridge \quad
\textsuperscript{2}University College London \quad
\textsuperscript{3}University of Bristol\\
*Corresponding author: jiayang.li.9@bristol.ac.uk
}
}
\maketitle
\raggedbottom
\begin{abstract}
Local learning reduces cross-layer training dependencies,
but an efficient accelerator must still retain operands,
accumulate gradients, and select quantization scales.
We present an FPGA system that trains a CNN and its
regional auxiliary classifiers using a shared integer matrix
engine. Compact errors are expanded on chip, gradients are
folded tile by tile, and historical scales permit direct
output conversion. At 200~MHz, the KV260 prototype achieves
76.57\% accuracy on CIFAR-10 and 91.85\% on SVHN at
approximately 4.20~W mean module power. Effective matrix
throughput is 31.22 and 31.57~GMAC/s, respectively.
Against the same integer method on Jetson Orin Nano,
measured training energy is 40.4\% and 42.1\% lower,
with $1.86\times$ and $1.82\times$ longer runtime;
measurements use the respective platform power domains.
Selected state checks and matching complete-run prediction
counts support functional agreement with the software reference.
\end{abstract}
\section{Introduction}

On-device training allows edge devices to learn from local
data and adapt models to individual users without uploading
private data. However, gradient computation and weight updates
add computation, while retaining forward inputs consumes
memory and recomputing them adds further work.
Reducing computation, intermediate storage, and data movement
while maintaining learning accuracy is therefore a key challenge
for training on resource-constrained edge devices.

Local learning and low-precision arithmetic address these
costs in complementary ways. By assigning an objective to
each layer or block, local learning enables updates without
waiting for errors from later layers and can shorten
activation lifetimes~\cite{local,ff,cwc}.
Low precision reduces storage per value; NITRO-D and
FF-INT8 demonstrate integer local training and its potential
on GPUs~\cite{nitro,ffint8}.
These algorithmic results, however, do not establish how
convolutional and local-learning operations should share
limited hardware resources, or how their intermediate data
should be stored and reused.

Prior FPGA accelerators for backpropagation share compute
resources across forward, backward, and weight-gradient
operations~\cite{shao,eftrain,venkataramanaiah}.
Our design provides local supervision through a trainable
classifier attached to each convolutional layer.
This removes cross-layer error propagation but introduces
the computations needed to train each classifier and generate
errors for its associated layer.
We map these computations and convolutional operations
to a shared integer matrix engine, avoiding a separate
matrix engine for the local classifiers.
Training completes one layer before advancing, so its saved
input storage can be reused after gradient computation.

We reduce the data that must be expanded, buffered, or
accumulated at once. Regional errors remain compact in DDR
and are expanded only for convolutional weight gradients.
Tile-wise folding bounds on-chip gradient storage, while
scales selected from earlier batches~\cite{dse,flexpoint}
allow direct INT32-to-INT8 output conversion after warmup.

The KV260 prototype completes CIFAR-10 and SVHN training
at 76.57\% and 91.85\% accuracy, using 19.62 and
32.01~kJ of measured module energy. It uses 40.4\% and
42.1\% less energy than the same integer method on Jetson,
with $1.86\times$ and $1.82\times$ longer runtime under
the measurement boundaries in Section~\ref{sec:evaluation}.
Controlled RTL experiments and storage accounting separately
characterize the cost of intermediate representations.

The contributions of this work are twofold:
\begin{contributions}
    \item \textbf{A shared architecture for integer local
    CNN training.}
    The FPGA trains both a CNN backbone and its local classifiers from random initialization using integer-only learning operations. Eight $8\times8$ INT8 compute units support convolutional and classifier computations, including score generation, errors for the associated convolutional layers, and weight gradients.

    \item \textbf{Reduced intermediate storage and memory
    traffic.}
    We combine compact regional-error storage with on-chip
    expansion, tile-wise gradient accumulation, and direct
    output conversion using historical scales.
    For the evaluated network at batch size 256, regional
    errors require 64--256~KiB per layer, compared with
    8--16~MiB for expanded INT8 spatial error maps.
    Tile accumulation limits on-chip accumulator capacity,
    while direct conversion avoids raw-output writes and
    reads in external memory after warmup.
\end{contributions}

\section{Integer Local Training}
\label{sec:method}

\subsection{CNN Architecture and Local Training}

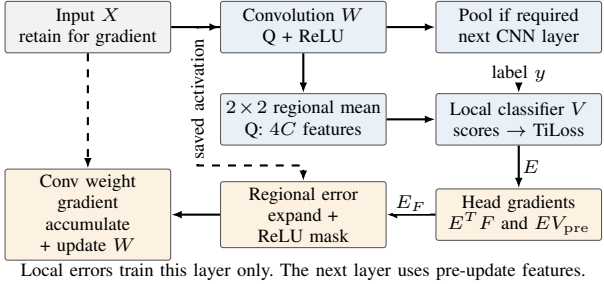
\begin{figure}[!t]
\centering
\begin{tikzpicture}[x=1cm,y=1cm,
 n/.style={draw=black!65,rounded corners=1pt,fill=compute,
 text width=2.05cm,minimum height=.68cm,align=center,
 font=\scriptsize,inner sep=2pt},
 a/.style={-{Latex[length=1.4mm]},thick},
 d/.style={a,dashed},t/.style={font=\scriptsize,fill=white,inner sep=1pt}]
\node[n,fill=storage] (x) at (1.15,0) {Input $X$\\retain for gradient};
\node[n] (conv) at (4,0) {Convolution $W$\\Q + ReLU};
\node[n] (next) at (6.85,0) {Pool if required\\next CNN layer};
\draw[a] (x)--(conv); \draw[a] (conv)--(next);
\node[n] (avg) at (4,-1.22) {$2\times2$ regional mean\\Q: $4C$ features};
\node[n] (head) at (6.85,-1.22) {Local classifier $V$\\scores $\to$ TiLoss};
\draw[a] (conv)--(avg); \draw[a] (avg)--(head);
\node[t] (label) at (6.85,-.62) {label $y$};
\draw[a] (label)--(head);
\node[n,fill=control] (vg) at (6.85,-2.48)
 {Head gradients\\$E^TF$ and $EV_{\rm pre}$};
\node[n,fill=control] (expand) at (4,-2.48)
 {Regional error\\expand + ReLU mask};
\node[n,fill=control] (wg) at (1.15,-2.48)
 {Conv weight gradient\\accumulate + update $W$};
\draw[a] (head)--node[t,right] {$E$}(vg);
\draw[a] (vg)--node[t,above] {$E_F$}(expand);
\draw[a] (expand)--(wg);
\draw[d] (x)--(wg);
\draw[d] (conv.west)--(2.6,0)--(2.6,-1.83)--(4,-1.83)--(expand.north);
\node[t,rotate=90] at (2.6,-.95) {saved activation};
\node[font=\scriptsize,align=center] at (4,-3.23)
 {Local errors train this layer only. The next layer uses pre-update features.};
\end{tikzpicture}
\caption{One CNN layer and its local supervision. Dashed
paths retain operands; the local head also updates $V$.
No convolutional input gradient is sent to earlier layers.}
\label{fig:local-training}
\end{figure}

The network contains a four-layer CNN backbone and a final
classifier (Fig.~\ref{fig:local-training}). Bias-free $3\times3$ standard and grouped
convolutions alternate~\cite{cwc}, each followed by
requantization (Q) and ReLU. Max pooling follows layers 2 and 4.

Following the quadrant-averaged auxiliary classifiers of
Belilovsky et al.~\cite{belilovsky}, each layer uses a trainable
local classifier on its quantized ReLU outputs before pooling.
Each channel is averaged over four equal spatial regions. The resulting $4C$ features
are quantized and mapped to $K$ class scores, requiring
$4CK$ weights rather than $HWCK$ for flattened features.

For $B$ samples, let $F\in\mathbb{Z}^{B\times4C}$ denote
the regional feature codes and
$V_{\mathrm{pre}}\in\mathbb{Z}^{K\times4C}$ the pre-update
classifier weights. The raw integer accumulations are
\begin{equation}
Z=FV_{\mathrm{pre}}^T,\qquad
G_V=E^TF,\qquad
E_F=EV_{\mathrm{pre}},
\label{eq:local-classifier}
\end{equation}
where $Z$ contains scores, $G_V$ classifier weight gradients,
and $E_F$ regional feature errors. Quantized scores and
labels enter TiLoss to produce classification errors $E$.

After requantization, each regional error is broadcast over
its region with the averaging factor $1/A$, where $A=HW/4$.
Power-of-two region sizes allow this factor to be applied
through an exponent adjustment. A ReLU mask suppresses
errors where the quantized activation is zero. These errors
train the associated convolutional layer without propagating
through earlier backbone layers.

A logical batch comprises the samples contributing to one
update. Training completes one layer before advancing,
and the next layer uses its pre-update forward outputs.
The final classifier uses flattened features and a batch-shared
dropout mask with keep probability $0.5$, doubling retained
values. Its errors update only this classifier. Evaluation
disables dropout and local classifiers and selects the largest
final INT32 score, breaking ties by the lowest class index.

\subsection{Integer Representation and Initialization}
\label{sec:integer-representation}

Each tensor represents values $q2^e$ using integer codes $q$
and a shared exponent $e$. Matrix operations use INT8 operands
and INT32 accumulators; the accumulator exponent is the sum
of the operand exponents. Conversion to INT8 divides every
code in the tensor by the same power of two:
\begin{equation}
q_i'=\operatorname{sat}_8
\left(\operatorname{RN}(q_i2^{-s})\right),
\qquad e'=e+s,\quad s\geq0,
\label{eq:conversion}
\end{equation}
where $\operatorname{RN}$ rounds to nearest with ties away
from zero, and $\operatorname{sat}_8$ clamps to $[-127,127]$.
The exponent adjustment compensates for the division before
rounding and saturation. The symmetric range permits
negation within INT8.

Regional features, training scores, and local feature errors
use $s=\max(0,b(m)-7)$, where $m=\max_i|q_i|$.
The range index $b(m)$ is the smallest
$b\in\{0,\ldots,30\}$ satisfying $m\leq T_b$,
or 31 if none exists; thus $b(0)=b(1)=0$.\footnote{
The thresholds are shared by the software and hardware:
$T_b=2^b$ for $0\leq b\leq20$; for $b=21,\ldots,30$,
the offsets added to $2^b$ are
$1,2,5,10,21,43,87,175,351,703$, respectively.}
Convolutional outputs use
Section~\ref{sec:historical-scales}.

The host tensor-rescales Xavier-normal samples
$W_{\mathrm{float}}$ by $127/m_W$ and rounds to nearest
with ties to even, where $m_W=\max|W_{\mathrm{float}}|>0$.
The weight exponent
$e_W=\lceil\log_2m_W\rceil-7$ remains fixed during training.
All-zero tensors are rejected.

\subsection{Historical Quantization Scales}
\label{sec:historical-scales}

Selecting a scale from current outputs delays conversion
until their range is known. Following historical-scaling
approaches~\cite{dse,flexpoint}, we select convolutional scales
from earlier batches. Our policy uses histogram percentiles
and a four-batch history window. The first eight logical
batches use their current maximum absolute output codes
to select the conversion shift under
Section~\ref{sec:integer-representation}, while collecting
historical candidates.

Each layer collects a histogram of raw INT32 code magnitudes:
a zero bin and nonzero bins
$(0,1],(1,2],\ldots,(2^{30},2^{31}]$.
For batch $t$ containing $N$ outputs, let $U_t$ be the upper
bound of the first bin reaching cumulative count
$\lceil999N/1000\rceil$. For $U_t>0$, the candidate exponent is
\begin{equation}
h_t=e_{\mathrm{raw},t}
+\min\{s\in\mathbb{Z}_{\geq0}:U_t\leq127\cdot2^s\},
\label{eq:history-candidate}
\end{equation}
where $e_{\mathrm{raw},t}$ is the accumulator exponent.
If the selected bin is zero, $h_t$ takes the largest available
candidate in the preceding four-batch window, or
$e_{\mathrm{raw},t}$ if history is empty.

After the first eight batches, conversion uses
\begin{equation}
\begin{aligned}
e_{\mathrm{target},t}&=\max_{i=1,\ldots,4}h_{t-i},\\
s_t&=\max(0,e_{\mathrm{target},t}-e_{\mathrm{raw},t}).
\end{aligned}
\label{eq:history-scale}
\end{equation}
The actual output exponent is
$e_{\mathrm{out},t}=e_{\mathrm{raw},t}+s_t$.
Conversion proceeds while current statistics are collected
for subsequent batches. Saturation handles values outside
the selected range; historical scales do not guarantee a
fixed clipping rate. Evaluation fixes each layer's target
to the maximum of its last four training candidates.

\subsection{Integer Classification Errors}
\label{sec:classification-errors}

We adapt the public TiLoss implementation of
NITI~\cite{niti} to generate INT8 surrogate errors from
quantized classifier scores and labels. We retain its
base-two exponential approximation, quadratic approximation,
and target-class error construction. For FPGA execution,
we implement the larger-exponent branch with lookup tables,
replace normalization divisions with a shared reciprocal,
and bound the small-exponent polynomial using a fixed scale.

For score exponents $e>-7$, the lookup tables supply
denominator contributions and scaled numerators $N_k$.
After summing the contributions to obtain $D$, we compute
one reciprocal with 28 fractional bits per sample:
\begin{equation}
R=\left\lfloor\frac{2^{28}}{D}\right\rfloor,\qquad
u_k=\left\lfloor\frac{N_kR}{2^{28}}\right\rfloor.
\label{eq:tiloss-normalization}
\end{equation}
All classes reuse $R$, replacing their normalization
divisions with integer multiplications and shifts.

For $e\leq-7$, NITI's public implementation represents
the quadratic approximation
$\exp(q_k2^e)\approx1+q_k2^e+q_k^22^{2e-1}$
as $2^{1-2e}+q_k2^{1-e}+q_k^2$.
Its scale grows as $e$ decreases. We retain this form
for $-9\leq e\leq-7$ and use a fixed $2^{15}$ scale
for $e\leq-10$:
\begin{equation}
a_k=2^{15}
+\operatorname{RN}(q_k2^{15+e})
+\operatorname{RN}(q_k^22^{14+2e}).
\label{eq:tiloss-polynomial}
\end{equation}
For INT8 score codes, this branch produces a 16-bit
unsigned polynomial value using bounded shifts.
Both polynomial forms supply $N_k=2^{11}a_k$ and
$D=\sum_ja_j$ to the same normalization path.

Non-target errors retain $u_k$, while the target-class
error subtracts $\sum_ju_j$ and receives a residual-based
rounding correction. Stochastic division by 16 and
saturation produce INT8 errors with exponent zero.
Backward computation does not differentiate rounding,
saturation, or scale selection.

\subsection{Gradient Accumulation and Weight Updates}
\label{sec:weight-updates}

Gradients are summed over each logical batch without
averaging. Eight-sample groups produce partial convolutional
gradients with matching exponents, which are accumulated
before updating weights. For the complete gradient tensor
$G$, GB5 normalization~\cite{niti} uses
\begin{equation}
s_G=\max(0,b(\max|G|)-5),\qquad
g=\operatorname{sat}_8
\left(\operatorname{RN}(G2^{-s_G})\right).
\label{eq:gb5}
\end{equation}
Gradients with $\max|G|\leq1$ are suppressed.
The five-bit magnitude scale is a target; rounding can
produce magnitude 32. Convolutional and local-classifier
weights follow the code-domain update
\begin{equation}
W^{+}=\operatorname{sat}_8(W-g),
\label{eq:local-update}
\end{equation}
without aligning the gradient exponent to the fixed weight
exponent.

To retain updates smaller than one INT8 weight step,
the final classifier uses a higher-resolution master
state. The INT16 state is initialized as
$M=16W$ and represents weights at exponent $e_W-4$.
After GB5, the gradient exponent is
$e_G=e_X+e_E+s_G$, where $e_X$ and $e_E$ are the input
and error exponents. With learning-rate shift $\lambda_t$
and $d=e_G-e_W+4-\lambda_t$,
\begin{equation}
\begin{aligned}
\Delta&=\operatorname{clip}_{[-2032,2032]}
\left(\operatorname{RN}(g2^d)\right),\\
M^{+}&=\operatorname{clip}_{[-2032,2032]}(M-\Delta),\\
W^{+}&=\operatorname{sat}_8
\left(\operatorname{RN}(M^{+}/16)\right).
\end{aligned}
\label{eq:master-update}
\end{equation}
Arithmetic is widened or saturation is determined before
overflow.

Scheduled freezing reduces later training work by skipping
selected layers' local classifiers, gradients, and updates,
while retaining forward computation and scale collection.
The final classifier remains trainable; Section~IV specifies
the schedule.

This procedure creates three hardware requirements:
(1) saved inputs and pre-update classifier weights must remain
available until their final gradient-related reads;
(2) GB5 updates require the complete batch-gradient range;
and (3) convolutional conversion requires an available scale.
Historical scales satisfy the third requirement without
waiting for current-output statistics.
Section~\ref{sec:architecture} describes the corresponding
storage, accumulation, and conversion architecture.

\section{Hardware Architecture}
\label{sec:architecture}

\begin{figure}[!t]
\centering
\begin{tikzpicture}[x=1cm,y=1cm,
 n/.style={draw=black!65,rounded corners=1pt,fill=compute,
 align=center,font=\scriptsize,inner sep=3pt},
 a/.style={-{Latex[length=1.4mm]},thick},
 bi/.style={{Latex[length=1.4mm]}-{Latex[length=1.4mm]},thick},
 t/.style={font=\scriptsize,fill=white,inner sep=1pt}]
\node[n,fill=control,text width=2.0cm,minimum height=.7cm]
 (ps) at (1.25,0) {PS controller\\batches, RNG, history};
\node[n,fill=storage,text width=4.7cm,minimum height=.7cm]
 (ddr) at (5.55,0) {DDR: parameters and training tensors\\saved inputs, compact errors, gradient buffers};
\draw[bi] (ps)--(ddr);
\draw[rounded corners=2pt,black!60] (.05,-.85) rectangle (8.05,-5.8);
\node[t,anchor=west] at (.2,-.85) {Programmable logic};
\node[n,fill=storage,text width=7.15cm,minimum height=.48cm]
 (axi) at (4.05,-1.45) {Single 128-bit AXI master / memory access};
\draw[bi] (ddr.south)--(5.55,-1.18);
\node[n,text width=1.8cm,minimum height=1.05cm]
 (supply) at (1.25,-2.78) {Operand buffers\\line / weight tiles\\broadcast};
\node[n,text width=2.15cm,minimum height=1.05cm]
 (engine) at (4.05,-2.78) {Shared matrix engine\\8 pods: $8\times8$ each\\INT8 $\to$ INT32};
\node[n,text width=1.65cm,minimum height=1.05cm]
 (collect) at (6.8,-2.78) {Result FIFOs\\collectors\\128-bit output};
\draw[bi] (1.25,-1.73)--(supply);
\draw[a] (supply)--(engine); \draw[a] (engine)--(collect);
\draw[a] (collect.north)--(6.8,-1.73);
\node[n,fill=control,text width=1.8cm,minimum height=1.05cm]
 (expand) at (1.25,-4.43) {Regional-error cache\\quadrant expansion\\activation mask};
\node[n,fill=control,text width=2.15cm,minimum height=1.05cm]
 (numerics) at (4.05,-4.43) {Numerical units\\TiLoss / GB5 / updates\\48-bit tile folding};
\node[n,fill=control,text width=1.65cm,minimum height=1.05cm]
 (out) at (6.8,-4.43) {Output units\\shift / ReLU / pool\\raw histogram};
\draw[a] (expand)--(supply);
\draw[a] (collect)--(out);
\draw[a] (collect.south)--(6.8,-3.53)--(4.05,-3.53)--(numerics.north);
\draw[thick] (.55,-5.48)--(7.5,-5.48);
\draw[bi] (expand.south)--(1.25,-5.48);
\draw[bi] (numerics.south)--(4.05,-5.48);
\draw[bi] (out.south)--(6.8,-5.48);
\draw[thick] (7.5,-5.48)--(7.8,-5.48)--(7.8,-1.45)--(7.65,-1.45);
\node[t] at (4,-5.48) {Shared memory access};
\draw[dashed,a] (ps.west)--(-.15,0)--(-.15,-2.05)--(4.05,-2.05)--(engine.north);
\node[t] at (3,-2.05) {control / selected operation};
\end{tikzpicture}
\caption{Hardware organization. Convolution and classifier
operations reuse the matrix engine. Regional errors expand
in operand supply; numerical and output units share DDR access.
Control is dashed; data connections are solid.}
\label{fig:architecture}
\end{figure}
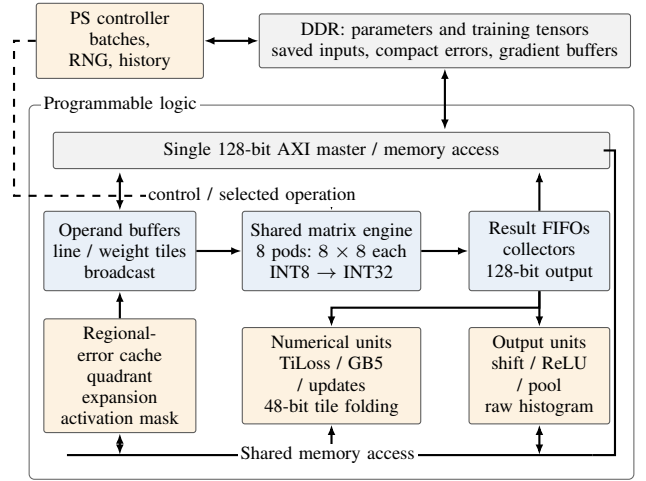

\subsection{Shared Engine and Operation Mapping}

Figure~\ref{fig:architecture} combines a shared integer matrix engine
with operand-supply, result-processing, and numerical units.
It addresses the requirements of Section~\ref{sec:method}
through operand retention, complete-gradient accumulation,
and output conversion with preselected scales.

The processing system (PS) prepares input batches and
random data, launches the programmable logic (PL), and
maintains scale history. DDR stores parameters and
intermediate tensors. The integrated training accelerator
accesses DDR through a single 128-bit AXI master.
Separate numerical units implement TiLoss, GB5 normalization,
and integer weight updates. Classifier scores supply the
error unit, while completed gradients supply the update
units. Random values for dropout and stochastic rounding
are supplied by the PS through DDR.

The matrix engine contains eight compute units, termed pods.
Each pod accumulates outer products of two eight-element
INT8 vectors into an $8\times8$ INT32 tile. Operations
reuse this arithmetic with different output and reduction
dimensions (Table~\ref{tab:engine-mappings}).
With all pods active and one reduction step accepted per
cycle, the engine performs 512 MACs per cycle, corresponding
to a theoretical peak of 102.4~GMAC/s at 200~MHz.

Local-classifier feature errors are computed using
pre-update weights, $E_F=EV_{\mathrm{pre}}$.
This supplies local supervision without convolutional
input-gradient propagation to preceding backbone layers.
Grouped convolutions restrict channel accesses to the
current group, and validity masks suppress lanes outside
tensor dimensions.

\begin{table}[t]
\centering
\caption{Shared-engine mappings.
$n$: sample; $p$: spatial position;
$c$: channel; $k$: class; $f$: feature.}
\label{tab:engine-mappings}
\footnotesize
\begin{tabular}{@{}lll@{}}
\hline
Operation & Output indices & Reduction \\
\hline
Conv forward
& $(p,c_{\mathrm{out}})$ & Kernel, $c_{\mathrm{in}}$ \\
Conv weight gradient$^{*}$
& $(c_{\mathrm{out}},c_{\mathrm{in}})$ & $n,p$ \\
Local-classifier forward
& $(n,k)$ & $f$ \\
Local feature error
& $(n,f)$ & $k$ \\
Classifier weight gradient
& $(k,f)$ & $n$ \\
Final-classifier forward
& $(n,k)$ & $f$ in four partitions \\
\hline
\end{tabular}

\smallskip
\parbox{\columnwidth}{\footnotesize
$^{*}$At a fixed kernel position.
For final-classifier forward computation, the collector
sums the four reduction partitions for each output tile.}
\end{table}

\subsection{Operand Supply and Result Collection}

At each reduction step, independent supply to eight pods
would require 16 eight-element vectors. Convolutional and
classifier weight gradients instead pair two error vectors
with four input vectors, requiring six distinct vectors.
Local-classifier forward uses four sample tiles and two class
tiles, also requiring six vectors. Feature-error computation
broadcasts one sample-error vector across eight feature tiles,
requiring nine vectors. Convolutional forward shares inputs
or weights according to the layer mapping.

For one sample tile, the final classifier partitions its
feature reduction four ways across two class tiles, assigning
all eight pods; the collector sums corresponding partial
results. Packed operand buffers provide convolution windows
and the classifier orientations in Table~\ref{tab:engine-mappings}.
Eight complete accumulator tiles require 128 transfers on
the result interface. Short reductions can therefore be limited
by collection, depending on overlap and backpressure.

\subsection{Regional-Error Expansion}

The convolutional weight-gradient supply path expands
compact regional errors when spatial error operands
are needed. This avoids storing a complete spatial error
tensor in DDR while preserving the local supervision
defined in Section~\ref{sec:method}.

The local classifier writes four INT8 error values per
channel and sample. The averaging factor $1/A$ is
represented by reducing their shared exponent by
$\log_2 A$ before convolutional weight-gradient computation.
Expansion therefore broadcasts integer codes without
performing a division at each spatial position.

For the current sample group and channel tile, the supply
logic loads the regional errors into an on-chip cache.
Each spatial position selects the error associated with
its quadrant. Saved quantized ReLU activations are read
from DDR, and their positive-value mask suppresses errors
at inactive positions. The expanded and masked errors
are buffered on chip and supplied to the matrix engine
alongside saved convolutional inputs.

This path avoids full spatial error-map writes and reads,
but still requires compact-error and activation accesses.
The current workspace retains two 16~MiB error-bank
allocations. The compact path stores regional errors
within this reserved space rather than materializing
dense spatial errors.

\subsection{Storage Reuse and Gradient Folding}

To satisfy the operand-lifetime requirement in
Section~\ref{sec:method}, storage is reused only after
each operand's final read. Convolutional inputs remain
available through weight-gradient computation, and
classifier source weights remain available through
feature-error computation.

Source parameter banks retain pre-update weights while
updates are written to destination banks. Training proceeds
layer by layer, allowing temporary storage to be reused
between phases. The next layer consumes the saved
pre-update forward output without recomputation.
The controller commits destination-bank selection after
the logical batch passes its completion checks.

To obtain the complete gradient and its range before
GB5 updates, the architecture folds partial gradients
one tile at a time. For $B=256$, eight-sample groups
produce 32 partial convolutional-gradient tensors.
These are stored in a dedicated DDR region until folding.
The completed gradient is written to a separate buffer.

Folding sums corresponding $8\times8$ tiles using a
48-bit accumulator with 384 bytes of logical storage.
Two 256-byte prefetch buffers alternate between loading
the next tile from DDR and supplying the current tile
for accumulation. Partial-gradient exponents are checked
before addition.

Each completed tile is checked against the INT32 range
and written to DDR. Its maximum absolute value contributes
to the complete-gradient range estimate. An out-of-range
result stops processing before updates. Once the full
range is known, the update pass rereads the gradient,
applies GB5 normalization, and updates convolutional
weights using \eqref{eq:local-update}.

For convolution 3, a resident 48-bit accumulator would
require 432~KiB of logical storage. Tile folding reduces
the accumulator to 384 bytes, with 512 bytes of prefetch
buffers, at the cost of DDR transfers. These figures
describe logical storage rather than physical BRAM
or URAM allocation.

\subsection{Output Conversion and Scale Collection}

To make the conversion scale available without waiting
for current-output statistics, the output path uses a
preselected historical shift. After the first eight
batches, each completed INT32 convolutional output reaches
two units: conversion applies this shift, while the
histogram observes the raw code before rounding
and saturation.

The histogram processes four INT32 values per 128-bit
word. Each lane alternates between two LUTRAM banks
on successive accepted words, separating concurrent
updates and spacing accesses to each bank.
The eight banks each contain 33 counters of 25 bits,
sufficient for the evaluated maximum of $2^{24}$ outputs
per layer and logical batch. HLS reports an initiation
interval of one cycle for the histogram loop.

After PL execution, the PS merges the bank counts,
checks the expected output count, and updates scale
history. Current statistics therefore do not delay
post-warmup conversion.

During the first eight batches, a 64~MiB DDR buffer
retains raw convolutional outputs. The PL collects
their extrema and subsequently converts the buffered
values within the same invocation. The concurrently
collected histogram initializes historical candidates.
The buffer remains reserved afterward: direct conversion
removes its raw-output writes and reads without reducing
the reserved capacity.

\section{Experimental Evaluation}
\label{sec:evaluation}

We evaluate complete FPGA training, effective throughput,
and storage costs, with Jetson experiments and published
FPGA results providing comparison.

\subsection{Experimental Setup}

Deployment uses seed 1, batch size $B=256$, eight-sample
gradient groups, channel widths 64/64/128/256, and
convolution groups 1/8/1/8.
CIFAR-10 uses 50,000 training images for 50 epochs;
SVHN uses 73,256 images for 60 epochs, omitting the
last training image to accommodate eight-sample grouping.
Incomplete batches are retained.
Accuracy uses the final model and official test set,
without augmentation or checkpoint selection.
Inputs use dataset normalization and integer exponents
$-5$ for CIFAR-10 and $-7$ for SVHN.
Layers freeze after 20/20/30/40 epochs; the final
classifier uses learning-rate shift 24 through epoch 40
and 25 thereafter.

FP32 backpropagation (BP) trains all layers throughout.
Local FP32 uses the same backbone, regional classifiers,
and freezing schedule as integer local training.
Both FP32 baselines use cross-entropy, Adam with
learning rate 0.001, and final dropout probability 0.5.

The KV260 operates at 200~MHz.
Jetson Orin Nano uses 25~W mode with a configured
GPU frequency of 918~MHz, PyTorch 2.8.0, CUDA 12.6,
and cuDNN 9.3.
Timing includes training execution but excludes data
preparation, initial loading, checkpoint writes, and testing.
Power samples, taken nominally every 5~ms on FPGA
and 200~ms on Jetson, are integrated over training
intervals without idle-power subtraction.
FPGA VCC\_SOM and Jetson VDD\_IN represent different
module power boundaries.

\subsection{FPGA Training and Throughput}
\label{sec:fpga-results}

\textbf{Complete training.}
The routed design meets 200~MHz timing with
$+0.003$~ns setup slack.
Table~\ref{tab:fpga} reports resources and training results.
The FPGA achieves 76.57\% accuracy on CIFAR-10 and
91.85\% on SVHN at approximately 4.20~W mean module
power.
Compared with the same integer-local configuration
on Jetson, measured training energy is 40.4\% and
42.1\% lower, while runtime is $1.86\times$ and
$1.82\times$ longer, respectively.

\begin{table}[!t]
\centering
\small
\setlength{\tabcolsep}{4pt}
\caption{KV260 implementation and complete training.
Energy is measured on VCC\_SOM; energy/sample counts
all training presentations across epochs.}
\label{tab:fpga}
\begin{tabular}{lrr}
\toprule
Resource & Count & Utilization (\%)\\
\midrule
LUT & 92,789 & 79.23\\
FF & 113,250 & 48.35\\
DSP & 587 & 47.04\\
BRAM (36-Kb equivalents) & 122.5 & 85.07\\
URAM & 20 & 31.25\\
\bottomrule
\end{tabular}

\par\smallskip

\begin{tabular}{lrr}
\toprule
Metric & CIFAR-10 & SVHN\\
\midrule
Accuracy (\%) & 76.57 & 91.85\\
Training time (s) & 4672.45 & 7626.85\\
Training energy (kJ) & 19.62 & 32.01\\
Mean module power (W) & 4.20 & 4.20\\
Energy/sample (mJ) & 7.85 & 7.28\\
\bottomrule
\end{tabular}
\end{table}

Selected first-batch and checkpoint-restoration tests
check parameter and random-generator states against
software references.
FPGA and Jetson integer runs also match cumulative
correct training predictions (1,645,036 and 3,796,354)
and final test counts (7,657/10,000 and 23,910/26,032).
These checks establish agreement for selected states
and aggregate outcomes, not every intermediate value.

\textbf{Effective throughput.}
PL execution accounts for 97.1\% and 96.8\% of
training time on CIFAR-10 and SVHN.
We divide nominal algorithmic matrix MACs by measured
PL time, counting convolutional forward computation,
active weight gradients and local-classifier operations,
and final-classifier forward and weight gradients.
Hardware-only padding and non-matrix operations are
excluded from the MAC count.

Table~\ref{tab:effective-throughput} shows the CIFAR-10
freezing phases.
Aggregate effective throughput is 31.22~GMAC/s
on CIFAR-10 and 31.57~GMAC/s on SVHN:
30.49\% and 30.83\% of the 102.4~GMAC/s peak.
PL time includes numerical processing, operand supply,
result collection, and memory service.

\begin{table}[!t]
\centering
\small
\setlength{\tabcolsep}{3pt}
\caption{Derived effective matrix throughput on CIFAR-10,
using algorithmic MAC counts and measured PL cycles.}
\label{tab:effective-throughput}
\begin{tabular}{lrrrr}
\toprule
Epochs & MMAC/sample & PL s/epoch & GMAC/s & Peak (\%)\\
\midrule
1--20  & 69.99 & 121.53 & 28.79 & 28.12\\
21--30 & 63.48 & 90.82  & 34.95 & 34.13\\
31--40 & 44.60 & 69.00  & 32.32 & 31.56\\
41--50 & 35.13 & 50.65  & 34.67 & 33.86\\
\bottomrule
\end{tabular}
\end{table}

The first two convolutional layers have forward
reductions of 36 and 72 steps, whereas eight complete
output tiles require 128 result transfers.
Short reductions therefore expose a result-bandwidth
constraint.
The complete-run counters do not separately attribute
time to memory stalls and result backpressure.

\textbf{Published FPGA context.}
Table~\ref{tab:prior-fpga} compares CIFAR-10 results
over 50 epochs.
The prior systems use six convolutional layers with
widths 16/16/32/32/64/64 and backpropagation,
whereas ours uses four wider, partly grouped layers
and local supervision.
Accuracy differences therefore include network and
training choices.

\begin{table}[!t]
\centering
\small
\setlength{\tabcolsep}{3pt}
\caption{CIFAR-10 FPGA training over 50 epochs.
Networks and power boundaries differ.}
\label{tab:prior-fpga}
\begin{tabular}{lrrr}
\toprule
Metric & EF-Train~\cite{eftrain}
       & Venk.~\cite{venkataramanaiah} & Ours\\
\midrule
Device & ZCU102 & Stratix 10 GX & KV260\\
Clock (MHz) & 100 & 240 & 200\\
Precision & FP32 & Fixed16 & Mixed integer\\
DSPs & 1,315 & 1,699 & 587\\
Accuracy (\%) & 64.82 & 73 & 76.57\\
Power (W) & 6.89$^\dagger$ & 20.64$^{\dagger *}$
          & 4.20$^\ddagger$\\
Energy/sample (mJ) & 14.33$^*$ & 7.43$^*$ & 7.85\\
\bottomrule
\end{tabular}

\par\smallskip
\begin{minipage}{\columnwidth}
\footnotesize
Ours: INT8 operands, INT32 matrix accumulation,
INT16 final master weights.
$^\dagger$Tool-estimated FPGA power;
$^\ddagger$measured module power;
$^*$derived.
EF-Train energy:
$6.89~\mathrm{W}\times2.08~\mathrm{ms}$.
Venk. power: sum of reported components; energy:
$20.64~\mathrm{W}\times18.01~\mathrm{s}/50{,}000$.
Different power boundaries preclude a normalized
energy ranking.
\end{minipage}
\end{table}

\subsection{Storage and Conversion Costs}

\textbf{Storage accounting.}
For $B=256$, regional errors occupy 64, 64, 128,
and 256~KiB across the four layers, versus
16, 16, 8, and 16~MiB for expanded INT8 spatial errors.
These $64$--$256\times$ reductions describe representation
capacity; traffic also depends on padding, repeated
reads, and activation accesses for masking.

For convolution 3, tile folding replaces a 432~KiB
resident 48-bit accumulator with a 384-byte accumulator
and 512-byte prefetch buffers.
The corresponding gradient-buffer payload is
18.5625~MiB per full batch for partial writes,
folding reads, complete-gradient writes, and update reads.
Folding trades external traffic for bounded on-chip
accumulation.

Reserved DDR workspace remains 161.22~MiB:
48~MiB features, 32~MiB error buffers,
64~MiB raw outputs, 9~MiB partial gradients,
and 8.22~MiB other state.
Compact errors use part of the reserved error space,
and the warmup buffer remains allocated.
These logical capacities and payload counts are
distinct from peak live storage and measured DDR traffic.

\textbf{Output conversion.}
Direct conversion avoids 448~MiB of raw-output
write/read payload per full batch after warmup.
A controlled RTL comparison uses identical inputs,
fixed shifts, and numerical processing, comparing
direct conversion with full-tensor staging and replay.
For 262,144 INT32 values, shift seven, ReLU enabled,
and no pooling, always-ready 128-bit interfaces
require 65,614 versus 131,157 cycles, a 50.0\%
reduction and 2~MiB less transfer payload.

A fixed throttling model gives 114,788 versus
316,886 cycles, a 63.8\% reduction.\footnote{
Input, output, scratch, and histogram service are
available on 4/5, 4/7, 2/4, and 3/5 of cycles,
with a 32-cycle scratch-read startup delay.}
The direct and buffered modules use 2,776 and
3,299 routed LUTs at a 200~MHz target and pass
192 C++ and 38 RTL reference checks.
These are isolated module results, excluding convolution,
production-layout conversion, and external scratch
storage/control.

\subsection{Software Accuracy and Policy Tradeoffs}

\textbf{Jetson comparison.}
Table~\ref{tab:jetson} shows that integer local reduces
energy relative to FP32 BP by 23.8\% on CIFAR-10
and 27.0\% on SVHN, with accuracy reductions of
3.30 and 0.53 percentage points.
Compared with Local FP32, it achieves higher accuracy
and 25.8\% lower peak CUDA allocation, but takes
$2.10\times$ and $2.25\times$ longer.
Local FP32 uses optimized GPU floating-point operators,
while the integer path implements custom numerical rules;
the results reflect both training and implementation
choices.

\begin{table}[!t]
\centering
\small
\setlength{\tabcolsep}{3pt}
\caption{Jetson results in 25~W mode with a configured
GPU frequency of 918~MHz.
Energy: VDD\_IN; memory: peak CUDA allocation.}
\label{tab:jetson}
\begin{tabular}{lrrrr}
\toprule
Method & Acc. (\%) & Time (s) & kJ & MiB\\
\midrule
\multicolumn{5}{c}{CIFAR-10, 50 epochs}\\
\midrule
FP32 BP & 79.87 & 2852.11 & 43.20 & 544.69\\
Local FP32 & 74.43 & 1198.08 & 17.62 & 401.18\\
Integer local & 76.57 & 2515.06 & 32.91 & 297.83\\
\midrule
\multicolumn{5}{c}{SVHN, 60 epochs}\\
\midrule
FP32 BP & 92.38 & 5022.51 & 75.67 & 544.69\\
Local FP32 & 91.22 & 1859.79 & 27.79 & 401.18\\
Integer local & 91.85 & 4183.28 & 55.25 & 297.83\\
\bottomrule
\end{tabular}
\end{table}

\textbf{Additional software results.}
With $B=128$ and three seeds, integer History achieves
$99.20\pm0.04$\%, $90.50\pm0.31$\%,
$77.00\pm0.65$\%, and $50.93\pm1.82$\%
on MNIST, Fashion-MNIST, CIFAR-10, and CIFAR-100,
respectively (mean $\pm$ sample standard deviation),
using 20/50/50/146 epochs.
The first three use widths 20/80/240/480.
The six-layer CIFAR-100 model uses 20-class coarse
supervision in its first four local heads and
100-class supervision thereafter.
These software configurations extend regional averaging
and class counts beyond the evaluated FPGA build.

\textbf{Policy study.}
A separate seed-1 study uses an earlier fixed numerical
baseline and the same deployment epoch budgets.
History and current-batch scaling differ by at most
0.13 percentage points in accuracy and less than 1\%
in runtime; History reduces peak CUDA allocation
from 345.57 to 297.83~MiB.
During testing, History fixes training-derived targets,
whereas Current uses unlabeled statistics from each
batch of up to 256 samples.
Scheduled freezing reduces training time by 28.4\%
on CIFAR-10 and 32.3\% on SVHN, with accuracy
increases of 6.38 and 2.03 percentage points.
These within-study results show that freezing affects
both workload and learning outcome.

\balance
\section{Conclusion}
The KV260 prototype demonstrates complete integer local CNN
training with shared convolutional and classifier arithmetic.
Compact errors, tile folding, and historical scales reduce
intermediate representation and buffering requirements.
Measured training uses approximately 4.20~W and achieves
31.22--31.57~GMAC/s effective matrix throughput. Relative
to the Jetson implementation, it consumes less measured
energy but takes longer under different module power
boundaries. The current evidence covers one FPGA seed,
logical storage savings, and isolated conversion tests.
Future work will measure board-level mechanism ablations
and build-matched memory and result-stall breakdowns.

\clearpage
\nobalance

\end{document}